\documentclass[submission,copyright,creativecommons]{eptcs}
\providecommand{\event}{NCMA 2024} % Name of the event you are submitting to
\usepackage{iftex}
\usepackage[colorinlistoftodos]{todonotes}
\usepackage{float}
\usepackage{amssymb}
\usepackage{amsmath}
\usepackage{xypic}
\newcommand{\qed}{\hspace*{\fill}$\Box$}
\newcommand{\states}{\mbox{\textrm{States}}}
\newcommand{\result}{\mbox{\textrm{Result}}}
\DeclareMathOperator{\oldmod}{mod}
\renewcommand{\mod}{\oldmod\:}
\def\card#1{\vert #1\vert}
\newtheorem{theorem}{Theorem}
\newtheorem{prop}[theorem]{Proposition}
\newtheorem{definition}{Definition}

\ifpdf
  \usepackage{underscore}         % Only needed if you use pdflatex.
  \usepackage[T1]{fontenc}        % Recommended with pdflatex
\else
  \usepackage{breakurl}           % Not needed if you use pdflatex only.
\fi

\title{Determinism in Multi-Soliton Automata}

\author{Henning Bordihn
\institute{Institut für Informatik und Computational Science\\ 
        Universität Potsdam\\Potsdam, Germany}
\email{henning@cs.uni-potsdam.de}
\and
Helena Schulz
\institute{Fakultät für Elektrotechnik und Informatik\\
    TU Berlin\\ Berlin, Germany}
\email{schulz-helena@gmx.de}}

\def\titlerunning{Determinism in Multi-Soliton Automata}
\def\authorrunning{H. Bordihn, H. Schulz}
\begin{document}
\maketitle

\begin{abstract}
Soliton automata are mathematical models of soliton switching in chemical molecules. Several concepts of determinism for soliton automata have been defined.
The concept of strong determinism has been investigated for the case in which only a single soliton can be present in a molecule. In the present paper, several different concepts of determinism are explored for the multi-soliton case. It is shown that the degree of non-determinism is a connected measure of descriptional complexity for multi-soliton automata. A characterization of the class of strongly deterministic multi-soliton automata is presented. Finally, the concept of perfect determinism, forming a natural extension of strong determinism, is introduced and considered for multi-soliton automata.
\end{abstract}

\section{Introduction}
\noindent
Soliton automata represent a model based on the switching behaviour of certain chemical molecules in which the bonds between (mainly carbon) atoms posses alternating weights. When some kind of disturbance is injected, it
travels through the molecule like a wave (or likewise a particle). The disturbance is called \emph{soliton} as it travels through the molecule "unhindered", without loss of energy and without interference.
The bonds between the molecule's atoms are changed along the path the soliton takes. 
This results in a different molecule. Taking the so obtained molecules as states, one is led to a system which
behaves like an automaton.

For a brief account of the history of solitons we refer to~\cite{Lomdahl:LosAlamos1984a} and~\cite[pp.18--19]{Lu:Soliton}. 
An extensive list of references regarding soliton computations and soliton automata
can be found in~\cite{BorJur:Multiwave}. The notion of soliton automata is encountered in~\cite{DassowJurg:Soliton}. In that paper also the concepts of determinism and strong determinism of soliton automata are considered and have been further investigated in~\cite{DassowJurg:Sol:SingExt,DassowJurg:Sol:OneCyc}. Strong determinism requires that, for every possible start and target atom, a soliton can take at most one path leading through the molecule. The main simplification of soliton automata as considered in~\cite{DassowJurg:Soliton,DassowJurg:Sol:SingExt,DassowJurg:Sol:OneCyc} is the assumption that only one single soliton can be present in a molecule at the same time. This restriction has been overcome in~\cite{BorJur:Multiwave} (and the subsequent paper~\cite{Koss:Bursts}), where multi-soliton automata have been taken into consideration in which more than one soliton can travel through a molecule simultaneously. Several different concepts of determinism for multi-soliton automata are defined in~\cite{Schulz:BSc:2023} and~\cite{BorSchu:Determinism}.

The present paper aims to continue this line of research. In the next section, the necessary notions related to soliton automata and the various concepts of determinism are given. We restrict ourselves to the case of multi-soliton automata since single-soliton automata as considered in~\cite{DassowJurg:Soliton} are  special cases of multi-soliton automata. In addition to deterministic and strongly deterministic soliton automata, also the concepts of perfect determinism and the degree of non-determinism are defined. Perfect determinism describes a natural concept that is somewhat "in between" determinism and strong determinism for soliton automata. The degree of non-determinism is a measure of descriptional complexity quantifying the amount of non-determinism of soliton automata. Section~\ref{sec:SolitonAutomata} concludes with the proof showing that the degree of non-determinism is connected with respect to soliton automata, that is, for every positive integer~$g$, there is a soliton automaton with degree of non-determinism~$g$.

Section~\ref{sec:Graphs} extends the notions of determinism to graphs underlying soliton automata (namely the graphs representing the bonding structure of the molecules under consideration, called soliton graphs). Similarly to the results known for the single-soliton case~\cite{DassowJurg:Soliton}, we give a characterization of soliton graphs always inducing strongly deterministic soliton automata. In~\cite{DassowJurg:Soliton} it is shown that a single-soliton automaton is strongly deterministic if and only if its underlying graph is a tree or a so-called chestnut (see Definition~\ref{def:chestnut}). For the multi-soliton case, we show that a soliton graph is strongly deterministic if and only if it is a tree. Moreover, we prove that there is a chestnut which is not even perfectly deterministic. 
The paper concludes with a few remarks on open research questions related to the results presented here.

\section{Soliton Automata}
\label{sec:SolitonAutomata}

\noindent
First, we introduce some notation and review some basic notions.
The sets of positive integers, of non-negative integers and of
integers are denoted by  $\mathbb{N}$, $\mathbb{N}_0$ and $\mathbb{Z}$, respectively.
We use standard notation for sets. We write $\vert S \rvert$ for the cardinality of
a set $S$. When no confusion is likely, we omit set brackets for singleton sets.

An {\it alphabet\/} is a finite non-empty set the elements of which are called {\it symbols.\/}
Let $\Sigma$ be an alphabet.
The set of all (finite) words over $\Sigma$, including the empty word~$\lambda$, is denoted
by $\Sigma^\ast$; let $\Sigma^+=\Sigma^\ast\setminus\{\lambda\}$.
The length $\text{lg}(w)$ of a word $w\in\Sigma^\ast$ is defined by
$$\text{lg}(w)=
\begin{cases}0,&\text{if $w=\lambda$,}\\
1+\text{lg}(v),&\text{if $w=av$ with $a\in\Sigma$ and $v\in\Sigma^\ast$}.\\
\end{cases}$$

A {\it semi-automaton\/} is a construct $\mathcal{A}=(Q,\Sigma,\tau)$
where $Q$ is a non-empty set, $\Sigma$ is an alphabet and $\tau: Q\times\Sigma\to 2^Q$
is a mapping. The elements of $Q$ are called states; $\Sigma$ is the input
alphabet of $\mathcal{A}$; $\tau$ is the transition function of $\mathcal{A}$.
In this paper, we assume that $Q$ is finite and that, for all $q\in Q$
and all $a\in\Sigma$, $\tau(q,a)\not=\emptyset$. Moreover, we drop the prefix ``semi-''
as we do not consider any other kind of automata.

Let $\mathcal{A}=(Q,\Sigma,\tau)$ be an automaton. The transition function $\tau$ is extended
to $2^Q\times\Sigma^\ast$ as follows: for $R\subseteq Q$ and $w\in\Sigma^\ast$, let
$$\tau(R,w)=\begin{cases}R,&\text{if $w=\lambda$,}\\
\tau\left(\bigcup_{q\in R}\tau(q,a),v\right),&\text{if $w=av$ with $a\in\Sigma$ and $v\in\Sigma^\ast$.}\\
\end{cases}$$
For $w\in\Sigma^\ast$, let $\tau_w$ be the mapping defined by
$\tau_w(R)=\tau(R,w)$ for all $R\subseteq Q$. Instead of $\tau_w(R)$ we often write
$R\tau_w$. 

The automaton $\mathcal{A}$ is said to be deterministic if $\vert \tau_a(q) \rvert=1$ for all $a\in\Sigma$ and
all $q\in Q$. In that case $\tau_a$ is considered as a mapping of $Q$ into $Q$, that
is as a transformation of $Q$ rather than of $2^Q$.
Inputs $u$ and $v$ of $\mathcal{A}$ are said to be equivalent if and only if $\tau_u=\tau_v$.

A {\it graph\/} is a pair $G=(N,E)$ with $N$ the set of {\it nodes\/}
and $E\subseteq N\times N$ the set of edges. We consider only finite undirected
graphs. An edge connecting nodes $n$ and $n'$ is given both as
$(n,n')$ and $(n',n)$. Therefore, we require that, for $n,n'\in N$,
$(n,n')\in E$ if and only if $(n',n)\in E$ and that these represent
the same edge. Thus, any two nodes can be connected by at most one
edge.
A {\it path\/} is a sequence of nodes $n_0, n_1, ..., n_k$ such that for $0 \leq i < k$ the pair $(n_i, n_{i+1}) \in E$.

A {\it weight function\/} for $G$ is a mapping $w:N\times N\to\mathbb{N}_0$
satisfying
$$w(n,n')=w(n',n)\begin{cases}{}=0,&\hbox{if }(n,n')\notin E\\
{}>0,&\hbox{if }(n,n')\in E.
\end{cases}$$
A {\it weighted graph\/} is a triple $(N,E,w)$ such that $(N,E)$
is a graph and  $w$ is a weight function.

For a node $n$, the set $V(n)=\{n'\mid (n,n')\in E\}$ is the {\it vicinity\/}
of $n$. The {\it degree\/} of $n$ is $d(n)=\card{V(n)}$,
and the {\it weight\/} of $n$ is $w(n)=\sum_{n'\in V(n)} w(n,n')$.
A node $n$ is said to be {\it isolated\/} if $d(n)=0$,
{\it exterior\/} if $d(n)=1$, and {\it interior\/}
if $d(n)>1$.

We now provide several definitions regarding soliton automata.

\begin{definition}[\cite{DassowJurg:Soliton}]\label{def:solGraph}
A \emph{soliton graph\/} is a weighted graph $G=(N,E,w)$ satisfying the following
conditions:
\begin{enumerate}
\item $N$ is the finite, non-empty set of nodes.
\item $E\subseteq N\times N$ is the set of undirected edges, such that $(n,n')\in E$ if and only if $(n',n)\in E$.
\item Every node $n\in N$ has the following properties:
\begin{enumerate}
\item $(n,n)\notin E$.
\item $1\leq d(n)\leq 3$.
\item $w(n)\in\{1,2\}$ if $n$ is exterior, and $w(n)=d(n)+1$
if $n$ is interior.
\end{enumerate}
\item Every component (maximal connected subgraph) of $G$ has at least one exterior node.
\end{enumerate}
\end{definition}

\begin{figure}[H]
	\centering
	\includegraphics[width=0.6\linewidth]{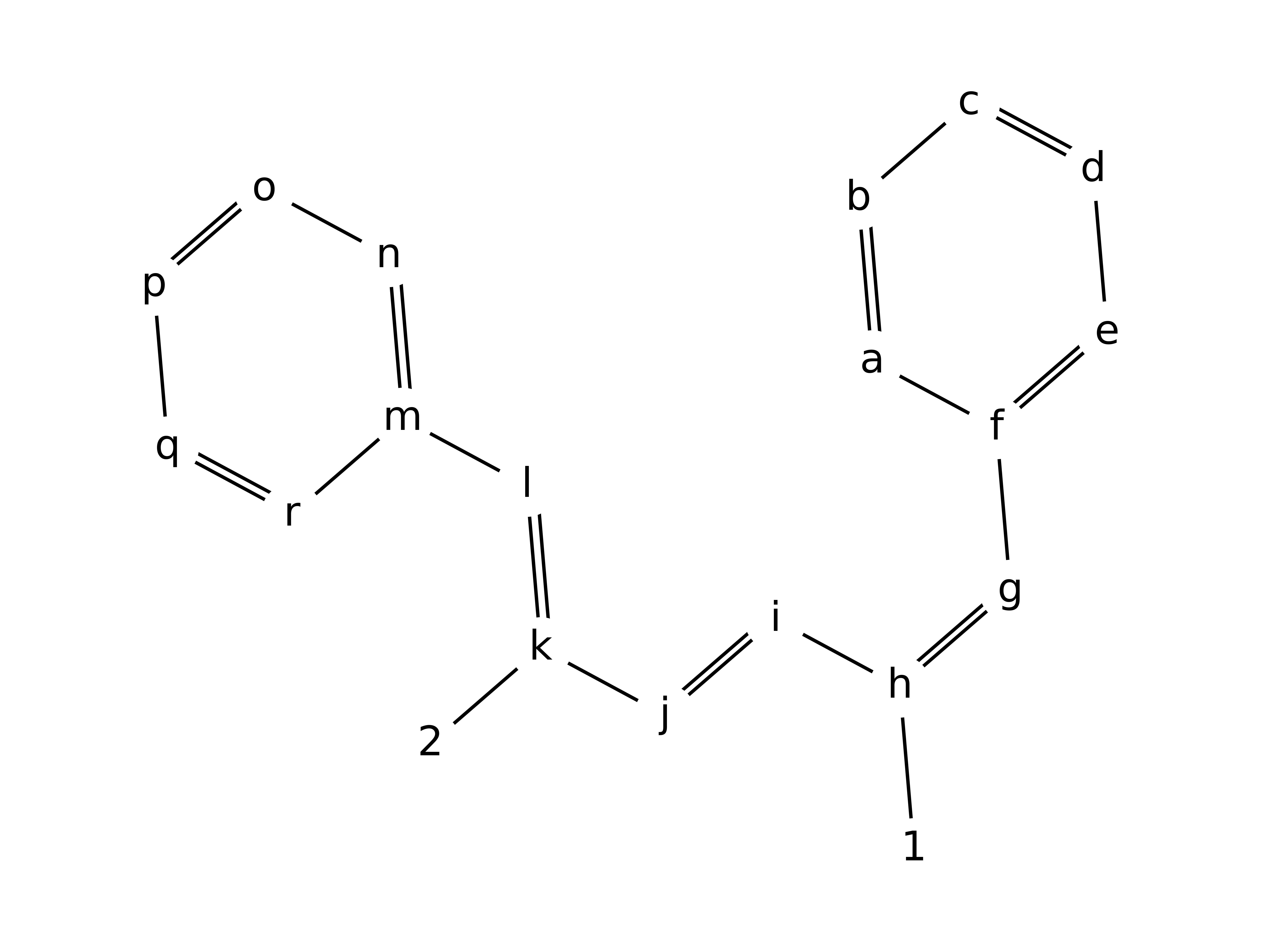}
	\caption{A soliton graph with two external nodes.}
	\label{fig:1}
\end{figure}

A soliton graph is an abstraction of a polyacetylene molecule.
Carbon atoms are represented as interior nodes, and connections to surrounding structures are represented as exterior nodes.
When drawing soliton graphs, we use letters for interior nodes and numbers for exterior nodes.
Just like in polyacetylene, only single and double bonds are allowed.
In the graph, single bonds are represented as edges with a weight of~$1$, and double bonds are represented as edges with a weight of~$2$.
We draw an edge of weight~$1$ as a simple line and an edge of weight~$2$ as two parallel lines.
The conditions regarding weight and degree imply that the two edges at a node of degree~$2$ must have different weights, and that, of the three edges meeting at a node of degree~$3$, two must have weight~$1$ and one must have weight~$2$.
An example of a soliton graph is depicted in Figure \ref{fig:1}.

In \cite{BorJur:Multiwave} it has been reasoned about properties of the abstract model.
The derived properties can be summarized as follows:
\begin{itemize}
	\item[(I)] One can insert and extract solitons at exterior nodes. 
    \item[(II)] Solitons move at a constant speed and have to move in every step. The speed is measured discretely as moving from one node to another one.
    \item[(III)] Solitons move at the same speed, which is why solitons cannot overtake each other on the same path.
    \item[(IV)] Solitons move over edges of alternating weights.
    \item[(V)] When a soliton travels along an edge of weight~$w$, the weight of the edge changes to $3 - w$.
    \item[(VI)] A soliton does not travel along the same edge twice in immediately consecutive steps.
    \item[(VII)] Multiple solitons cannot travel along the same edge in the same step.
\end{itemize}

\begin{definition}[Bursts of Inputs \cite{BorJur:Multiwave}]\label{def:burst}
Let~$S$ be a finite non-empty set not containing the symbols~$\|$ and~$\bot$.
Moreover, let $S\cap \mathbb{N}_0 = \emptyset$.

A \textit{burst over}~$S$ is a word of the form
$$s_1\|_{k_1}s_2\|_{k_2}\cdots s_{m-1}\|_{k_{m-1}}s_m\bot$$
with the following properties:
\begin{enumerate}
\item $m\in \mathbb{N}$;
\item $s_1,s_2,\ldots,s_m\in S$;
\item $k_1,k_2,\ldots,k_{m-1}\in \mathbb{N}_0$;
\end{enumerate}
The \emph{length\/} of such a burst is~$m$.

For $m\in \mathbb{N}$, let $\mathcal{B}_m(S)$ be the set of all bursts of length~$m$ over~$S$.
Let 
\[
\mathcal{B}_{\leq m}(S)=\bigcup_{i=1}^m \mathcal{B}_i(S) \text{ and } \mathcal{B}(S)=\bigcup_{i\geq1} \mathcal{B}_i(S).
\]
\end{definition}

Let~$G$ be a soliton graph, let~$X$ be the set of its exterior nodes and $S = X \times X$.
Then any set $B \subseteq \mathcal{B}(S)$ is called a set of bursts for~$G$.
The pair $s_i \in S$ contains the two nodes the $i$th soliton enters and leaves the graph through, respectively.
A burst of the form $s_1\|_{k_1}s_2\|_{k_2}\cdots s_{m-1}\|_{k_{m-1}}s_m\bot$ is to be interpreted as follows. If the burst is initiated at time $t$,
the symbol $s_1$ is input at time $t$; $s_2$ is input at time $t+{k_1}$; and,
in general, $s_j$ is input at time $t+\sum_{i=1}^{j-1}k_i$. Here the empty sum is defined to be 0.
The symbol $\bot$ indicates that the input process pauses until the system has stabilized.

\begin{definition}[Position Map \cite{BorJur:Multiwave}]\label{def:posMap}
For $m\in \mathbb{N}$, let $\mathfrak m=\{1,2,\ldots,m\}$.
Further, let $G=(N,E,w)$ be a soliton graph such that $N\cap \mathbb{N}_0=\emptyset$.
A \emph{position map\/} for $m$ is a mapping
of $\mathfrak m$ into $N\cup \mathbb{N}_0$.
\end{definition}

If~$\pi$ is a position map for~$m$, then $\pi(i)$ indicates
at which node the $i$\/th soliton is or how many steps are still
required until it will enter the graph. Thus $\pi(i)=1$ means that the ith
soliton will enter the graph in the next step. $\pi(i)=n$ with $n\in N$ means that the soliton
is at node~$n$. $\pi(i)=0$ means, by definition, that the $i$th soliton has left the graph.

\begin{definition}[Initial Position Map for a Burst \cite{BorJur:Multiwave}]\label{def:iniPosMap}
Let
$$b=(n_1,n_1')\|_{k_1}(n_2,n'_2)\|_{k_2}\cdots
(n_m,n'_m)\bot$$
be a burst of length $m$.
The \emph{initial position map $\pi_b$ for\/} $b$ is defined as follows:
Let $r$ be minimal such that $k_1=k_2=\cdots k_r=0$ and $k_{r+1}>0$ or $r=m-1$.
Then
$$\pi_b(i)=\begin{cases}n_i,&\text{if $1\leq i\leq r+1$,}\\
k_{r+1},&\text{if $i=r+2$,}\\
\pi_b(i-1)+k_{i-1},&\text{if $i>r+2$.}\\
\end{cases}$$
\end{definition}

For example, let
$$b=(n_1,n'_1)\|_0(n_2,n'_2)\|_3(n_3,n'_3)\|_1(n_4,n'_4)\|_0(n_5,n'_5)\bot$$
be a burst. Then $\pi_b$ is given by the following table:
\begingroup\offinterlineskip
\def\TKS{\omit&height2true pt&\omit&height2true pt&\omit&height2true pt&%
\omit&height2true pt&\omit&height2true pt&\omit&height2true pt\cr}
\def\LN{\noalign{\hrule}}
$$\vbox{%
\halign{\strut#\hfil\quad&\vrule#&&\hfil\quad\strut#\quad\hfil&\vrule#\cr
Soliton $i$&&1&&2&&3&&4&&5&\cr
\TKS
\LN
\TKS
Position $\pi_b(i)$&&$n_1$&&$n_2$&&3&&4&&4&\cr
}}$$
\endgroup

This means that the first two solitons start at node~$n_1$ and~$n_2$, respectively.
The other solitons have to wait for~$3$ or~$4$ time steps.

\begin{definition}[Final Position Map \cite{BorJur:Multiwave}]\label{def:finPosMap}
A position map $\pi$ for $m$ is said to be \emph{final\/}
if $\pi(i)=0$ for all $i\in\mathfrak m$.
\end{definition}

The processing of a burst starts with its initial position map and ends with a final position map corresponding in terms of the number of solitons.
Small intermediate steps occur leading from the initial position map to the final position map.
A burst is successful if and only if all its solitons have left the soliton graph after a finite amount of time.

\begin{definition}[Potential Successor Map \cite{BorJur:Multiwave}]\label{def:potSucMap}
Let $G$ be a soliton graph.
Let $m\in \mathbb{N}$, and let
$\pi$ and $\pi'$ be  position maps for $m$. Let
$$b=(n_1,n_1')\|_{k_1}(n_2,n'_2)\|_{k_2}\cdots
(n_m,n'_m)\bot$$
be a burst of length $m$.

The map $\pi'$ is a \emph{potential (direct) successor} of $\pi$ (with respect to $b$), if and only if
$$\pi'(i)=\begin{cases}\pi(i)-1,&\text{if $\pi(i)\in \mathbb{N}_0$ and $\pi(i)>1$,}\\
n_i,&\text{if $\pi(i)\in \mathbb{N}_0$ and $\pi(i)=1$,}\\
n,&\text{if $\pi(i)\in N$, $\pi(i)\not=n_i'$, $n\in N$, and $\bigl(\pi(i),n\bigr)\in E$,}\\
0,&\text{if $\pi(i)=n_i'$ or if $\pi(i)=0$.}\\
\end{cases}$$
for $i=1,2,\ldots,m$.
\end{definition}

This ensures that the waiting times of the solitons are reduced in every step,  solitons enter the graph at the right node, they have to use an edge in order to reach the next node, and that~$0$ is a value in the position map if the corresponding soliton reached the exterior node it is supposed to leave the graph through.

\begin{definition}[Configuration and Configuration Trail \cite{BorJur:Multiwave}]\label{def:configTrail}
Let $G=(N,E,w)$ be a soliton graph.
Let $m\in \mathbb{N}$, and let
$$b=(n_1,n_1')\|_{k_1}(n_2,n'_2)\|_{k_2}\cdots
(n_m,n'_m)\bot$$
be a burst of length $m$.
\begin{enumerate}
\item A \emph{configuration} (for $b$) is a pair $(G',\pi)$ such that
$G'=(N,E,w')$ is a weighted graph with weights in $\{1,2\}$ and
$\pi$ is a position map for $m$.

\item A \emph{configuration trail} for $G$ and $b$ is a finite sequence
$$(G_0,\pi_0), (G_1,\pi_1),\ldots$$
of configurations for $b$ with the following  properties.
\begin{enumerate}
\item $G_0=G$, and $\pi_0$ is the initial position map for $b$.

\item
\label{Cond2b} $\pi_1$ is a potential successor of $\pi_0$ such that $\pi_0(i)\in N$
implies $\pi_1(i)\in N$ for all $i\in\mathfrak m$.\\
$G_1=(N,E,w_1)$ is obtained from $G_0=(N,E,w_0)$ by changing the weights of some edges
as follows:
If $\pi_0(i)\in N$, then
$$w_1\bigl(\pi_0(i),\pi_1(i)\bigr)=
w_1\bigl(\pi_1(i),\pi_0(i)\bigr)=
3-w_0\bigl(\pi_0(i),\pi_1(i)\bigr).$$
For all other edges the weights remain unchanged.

\item
\label{Cond2c}
Let $j>1$. The sequence
$$(G_0,\pi_0), (G_1,\pi_1),\ldots, (G_j,\pi_j)$$
is a configuration trail, if and only if
$$(G_0,\pi_0), (G_1,\pi_1),\ldots, (G_{j-1},\pi_{j-1})$$
is a configuration trail such that
$\pi_{j-1}$ is not final, $G_j = (N,E,w_j)$, and the following conditions are satisfied (for all $i\in\mathfrak m$):
\begin{enumerate}
\item $\pi_{j}$ is a potential successor of $\pi_{j-1}$.
\item
\label{Cond2cii}
If $\pi_{j-1}(i)\in N$ is exterior and $\pi_{j-2}(i)=1$, then $\pi_j(i)\in N$.
\item
\label{Cond2ciii}
If $\pi_{j-1}(i)\in N$ is exterior and
equal to $n_i'$, and if $\pi_{j-2}(i)\in N$, then $\pi_j(i)=0$.
\item If $\pi_{j-1}(i)\in N$ is interior and  $\pi_{j-2}(i)\in N$, then
$$w_{j-2}\bigl(\pi_{j-2}(i),\pi_{j-1}(i)\bigr)\not=
w_{j-1}\bigl(\pi_{j-1}(i),\pi_{j}(i)\bigr).$$
\item If $\pi_j(i)\not=0$, then
$\pi_j(i)\not=\pi_{j-1}(i)$
and
$\pi_j(i)\not=\pi_{j-2}(i)$.

\item $G_j$ is obtained from $G_{j-1}$ by changing the weights of some
edges as follows:\\
If $\bigl(\pi_{j-1}(i),\pi_{j}(i)\bigr)\in E$,
then
$$w_j\bigl(\pi_{j-1}(i),\pi_{j}(i)\bigr)=
w_j\bigl(\pi_{j}(i),\pi_{j-1}(i)\bigr)=
3-w_{j-1}\bigl(\pi_{j-1}(i),\pi_{j}(i)\bigr).$$
All other weights remain unchanged.

\end{enumerate}

\end{enumerate}

\item A configuration trail is \emph{legal}, if
it satisfies the following conditions for all $j\geq 1$:
\begin{enumerate}
\item If $\pi_{j-1}(i)$ and $\pi_{j-1}(i')$ are
nodes and $\pi_{j-1}(i)=\pi_{j-1}(i')$ for some distinct $i$ and $i'$,\\
then $\pi_j(i)\not=\pi_j(i')$.
\item If $\pi_{j-1}(i)$ and $\pi_{j-1}(i')$ are nodes
with $\bigl(\pi_{j-1}(i),\pi_{j-1}(i')\bigr)\in E$,
then
$\pi_j(i)\not=\pi_{j-1}(i')$ \\or
$\pi_j(i')\not=\pi_{j-1}(i)$.
\end{enumerate}
\item A configuration trail
$$(G_0,\pi_0), (G_1,\pi_1),\ldots, (G_j,\pi_j)$$
is \emph{partial} if $\pi_j$ is not final. Otherwise,
it is \emph{total}.

\end{enumerate}
\end{definition}

A configuration defines the weights of the current graph and the positions of the solitons for a certain time step.
Note that the graph in a configuration need not be a soliton graph.
It represents the situation when all solitons have reached the "next" nodes on their ways.
Consider, for example, the soliton graph in Figure~\ref{fig:1}.
If a soliton entered the graph at node~$1$ and has reached node~$h$, the weight of edge~$(1,h)$ has changed to~$2$; thus $w(h) = 5$.

The conditions above ensure that all solitons behave exactly as defined in the rules concerning soliton movements.
A consequence of the condition that no two solitons can traverse the same edge at the same time is that they also cannot enter the same exterior node at the same time.
This holds true both for exterior nodes used as entry points and those used as exit points.
Two solitons can be at an interior node simultaneously, but must leave it on different edges.
Moreover, they cannot simply swap places.

\begin{definition}[Soliton Path]
Let $G=(N,E,w)$ be a soliton graph.
Let $m\in \mathbb{N}$, let
$$b=(n_1,n_1')\|_{k_1}(n_2,n'_2)\|_{k_2}\cdots
(n_m,n'_m)\bot$$
be a burst of length $m$, and let
$C = (G_0,\pi_0), (G_1,\pi_1),\ldots, (G_j,\pi_j)$
be a configuration trail for~$G$ and~$b$, $j\ge 0$.
For every $i\in\mathfrak m$, let $\ell$ be the smallest and $r$ be the largest number, $0 \leq \ell \leq r \leq j$ such that $\pi_\ell(i) \in N$ and $\pi_r(i) \in N$.
The path
$$\pi_\ell(i),\pi_{\ell+1}(i),\ldots,\pi_r(i)$$
is the soliton path of soliton $i$ in $C$.
For $\ell\le h < r$, the edge $(\pi_h(i),\pi_{h+1}(i))$ is said to be \emph{used} by soliton~$i$ in~$C$.
\end{definition}

\begin{definition}[Result of a Burst \cite{BorJur:Multiwave}]\label{def:result}
Let~$G$ be a soliton graph and let~$b$ be a burst.
The \emph{result of burst~$b$ on~$G$}
is the set
$\result(G,b)$
of weighted graphs $G'$ such that there is a total legal
configuration trail for~$G$ and~$b$ transforming~$G$ into~$G'$.
\end{definition}

Every element of $\result(G,b)$ is again a soliton graph.

Let $B \subseteq \mathcal{B}(X \times X)$ be a set of bursts.
Let
\[
\result(G, B) = \bigcup_{b \in B} \result(G,b) \text{.}
\]
For $i \in \mathbb{N}_0$, let
\[
    \result^i(G, B) =
    \begin{cases}
    G,& \text{if } i = 0 \text{, and}\\
    \result(\result^{i-1}(G, B), B), & \text{if } i > 0
    \end{cases}
\]
and
\[
\result^*(G, B) = \bigcup_{i \geq 0} \result^i(G,B) \text{.}
\]

We can use the resulting soliton graphs we obtain by traversing total legal configuration trails as states of an automaton.
Such an automaton is induced by an underlying soliton graph and a set of bursts.

\begin{definition}[Multi-Soliton Automaton \cite{BorJur:Multiwave}]\label{def:MultSolAuto}
Let $G$ be a soliton graph with set~$X$ of exterior nodes. Let $B \subseteq \mathcal{B}(X \times X)$ be a set of bursts.
Let
$$\states(G,B)=\result^*(G,B).$$
The \emph{$B$-soliton automaton\/} of $G$ is the finite automaton $\mathcal{A}_B(G)$ with inputs $b\in B$, state set $\states(G,B)$ and non-deterministic transition function
$$\tau(G',b)=\begin{cases}\result(G',b),&\text{if $\result(G',b)\not=\emptyset$,}\\
\{G'\},&\text{otherwise,}\\
\end{cases}$$
for $G'\in \states(G,B)$ and $b\in B$.
\end{definition}

Note that $\states(G,B)$ is bounded, as the set of vertices and the set of edges do not change, only the weights do. 
Therefore, there is a finite set~$B$ of bursts
such that $\states(G,B)=\states(G,B')$ for all sets~$B'$ of bursts
with $B\subseteq B'$.
If there is no risk of confusion, a $B$-soliton automaton will be called multi-soliton automaton or simply soliton automaton.

Now we define different kinds of determinism for soliton automata.

\begin{definition}[Determinism \cite{BorSchu:Determinism}]\label{def:det}
Let $G = (N, E, w)$ be a soliton graph and let~$B$ be a set of bursts for~$G$. 

$\mathcal{A}_B(G)$ is called
\begin{itemize}
    \item[(I)] \emph{deterministic}, if $\vert \result(G', b) \rvert = 1$ for all $G' \in \states(G, B)$ and all $b \in B$.
    \item[(II)] \emph{strongly deterministic}, if for all $G' \in \states(G, B)$ and $b \in B$, there is at most one total legal configuration trail for~$G'$ and~$b$.
\end{itemize}
\end{definition}

A soliton automaton is deterministic if there is exactly one successor state for each state in the set $\states(G, B)$ and each burst in~$B$.
Strong determinism is an even stronger constraint, as it also restraints in how many ways it is possible to transition from one state into another.
An automaton has this kind of determinism if there is at most one total legal configuration trail for each state in $\states(G, B)$ and each burst in~$B$.

By definition, all total legal configuration trails are considered as transitions between the automaton's states.
There are, however, cases in which an infinite number of configuration trails are possible for a state and a burst.
For example, a soliton can get into a situation where it has the possibility to traverse a cycle infinitely many times.
Since configuration trails for those kinds of situations contain "unnecessary" repetitions, we aim to classify configuration trails into two categories: perfect and imperfect.
In order to determine when a configuration trail becomes imperfect, we search for equivalent configurations in it.
Therefore, we first need to describe when two configurations are called equivalent.

\begin{definition}[\cite{BorSchu:Determinism}] \label{def:sc}
Let~$G$ be a soliton graph.
Let $C = (G_0, \pi_0), (G_1, \pi_1), ..., (G_i, \pi_i)$ be a partial configuration trail that is not a total configuration trail.
The set of possible successor position maps, denoted as $SC(C)$, is the set containing all $\pi_{i+1}$, such that $(G_0, \pi_0), (G_1, \pi_1), ..., (G_i, \pi_i), (G_{i+1}, \pi_{i+1})$ is a configuration trail.
\end{definition}

\begin{definition} [Successor-Equivalence \cite{BorSchu:Determinism}] \label{def:succEquiv}
Let $(G_0, \pi_0), (G_1, \pi_1), ..., (G_k, \pi_k)$ be a configuration trail.
\linebreak
For integers $i$ and $j$ with $0 \leq i \leq k$ and $0 \leq j \leq k$, let $C = (G_0, \pi_0), (G_1, \pi_1), ..., (G_i, \pi_i)$ and let $C' = (G_0, \pi_0), (G_1, \pi_1), ..., (G_j, \pi_j)$.
The configurations $(G_i, \pi_i)$ and $(G_j, \pi_j)$ are called \textit{successor-equivalent}, if $(G_i, \pi_i) = (G_j, \pi_j)$ and $SC(C) = SC(C')$.
This property is written as $(G_i, \pi_i) \equiv_{SC} (G_j, \pi_j)$.
\end{definition}

\begin{definition} [(Im)Perfect Configuration Trail \cite{BorSchu:Determinism}] \label{def:imperfect}
Let~$G$ be a soliton graph and let~$b$ be a burst.
Let $C = (G_0, \pi_0), (G_1, \pi_1), ..., (G_k, \pi_k)$ be a configuration trail with $G_0 = G$ and $\pi_0 = \pi_b$ (the initial position map for $b$).
$C$ is called \textit{imperfect}, if at least two configurations $(G_i, \pi_i)$ and $(G_j, \pi_j)$ exist, $0 \leq i < j \leq k$, where $(G_i, \pi_i) \equiv_{SC} (G_j, \pi_j)$.
Otherwise, $C$ is called \textit{perfect}.
\end{definition}

A configuration trail is perfect, if there are no two occurrences of successor-equivalent configurations in it.
In the other case, we define it as imperfect, because then it would contain unnecessary steps. 
Consider the case of a configuration trail with two successor-equivalent configurations $(G_i, \pi_i)$ and $(G_j, \pi_j)$.
We could make the exact same next moves after time step~$i$ and~$j$, so we might as well cut out all the configurations from $(G_{i+1}, \pi_{i+1})$ until $(G_j, \pi_j)$.
In \cite{BorSchu:Determinism} it is shown that considering only perfect configuration trails in the construction of a soliton automaton does not change its set of states. 
Also, if an imperfect configuration trail exists in a soliton automaton it can not be strongly deterministic.

\begin{figure}[th]
	\centering
	\includegraphics[width=1.0\linewidth]{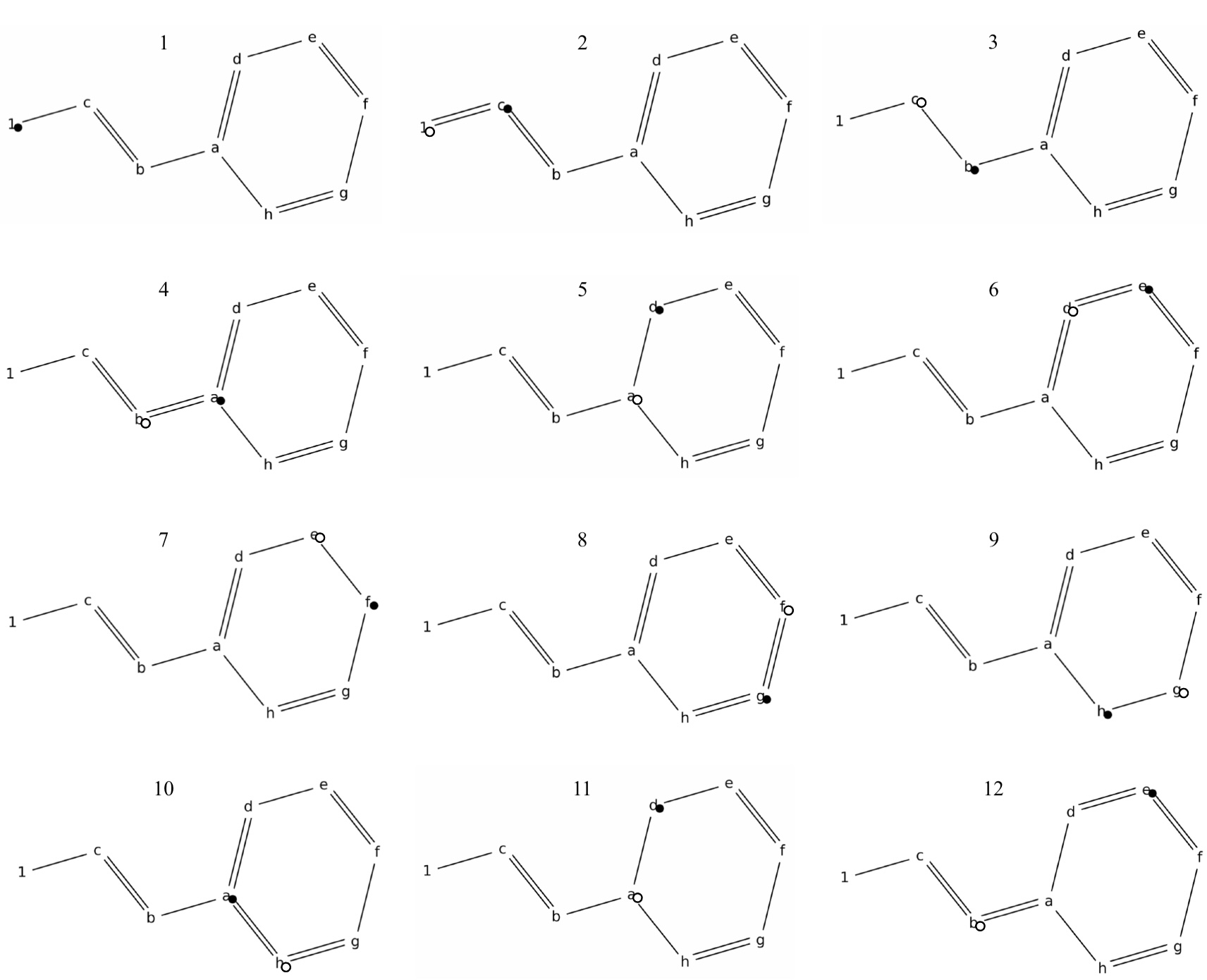}
	\caption{Part of a configuration trail for the burst $(1,1)\|_1(1,1)\bot$. The first soliton is depicted as a black pebble, while the second one is depicted as a white pebble.}
	\label{fig:2}
\end{figure}

Let~$G$ be the soliton graph from configuration~$1$ in Figure \ref{fig:2} and let~$B = \{(1,1)\bot\}$. 
In \cite{DassowJurg:Soliton} it is shown that the $B$-soliton automaton $\mathcal{A}_B(G)$ is strongly deterministic.
On the other hand, by using the set of bursts $B' = \{(1,1)\|_1(1,1)\bot\}$ the automaton $\mathcal{A}_{B'}(G)$ is not strongly deterministic. 
This is due to the white soliton having two possible successor positions in configuration~$11$.
It could move to node~$b$, like in configuration~$12$, or it could move to node~$d$, resulting in both solitons staying inside the cycle.
Eventually, this configuration trail would lead to a configuration successor-equivalent to configuration~$11$ and can therefore be classified as imperfect.
However, for this graph and this burst we cannot find any perfect total legal configuration trails, except from the trail continued from configuration~$12$.
In order to further discriminate such situations we introduce the following property.

\begin{definition}[Perfect Determinism]\label{def:perfectDet}
Let $G = (N, E, w)$ be a soliton graph and let~$B$ be a set of bursts for~$G$. $\mathcal{A}_B(G)$ is called \emph{perfectly deterministic}, if for all $G' \in \states(G, B)$ and $b \in B$, there is at most one perfect total legal configuration trail for~$G'$ and~$b$.
\end{definition}

The automaton $\mathcal{A}_{B'}(G)$ from our example is perfectly deterministic, but not strongly deterministic.
Hence, perfect determinism lies "in between" determinism and strong determinism.

Distinct soliton automata that are \emph{not} deterministic may be different "distances" away from fulfilling the determinism property.
We now formulate a measure of descriptional complexity that quantifies this distance.

\begin{definition}[Degree of Non-Determinism \cite{BorSchu:Determinism}]
Let $G = (N, E, w)$ be a soliton graph and let $B$ be a set of bursts for $G$.
The \emph{degree of non-determinism} of $\mathcal{A}_B(G)$ is the smallest integer $g \geq 1$, such that $\vert \result(G', b) \rvert \leq g$ for all $G' \in \states(G, B)$ and all $b \in B$.
\end{definition}

\begin{theorem}
    The degree of non-determinism is a \emph{connected} measure of descriptional complexity, that is, for every positive integer~$g$, there is a soliton automaton~$A_g$  such that its degree of non-determinism is~$g$.
\end{theorem}

\begin{figure}[hb]
  \begin{minipage}{0.18\textwidth}\tiny
    \[\xymatrix{ %1
        1 \ar@{-}[r] & n_1 \ar@{=}[r]     & n_2 \ar@{-}[ld] \\
                     &     v_1 \ar@{=}[d] &               \\
				       &     2             & 
}\]
  \end{minipage}\hfill
  \begin{minipage}{.28\textwidth}\tiny
    \[\xymatrix{ %2
       1 \ar@{-}[r] & n_1 \ar@{=}[r]     & n_2 \ar@{-}[dr] \ar@{-}[r] & n_3 \ar@{=}[r] & n_4 \ar@{-}[dl]\\
                    &                  &                          & v_1 \ar@{=}[d] &              \\
			      	&                  &                          & 2              &
}\]
  \end{minipage}\hfill
  \begin{minipage}{.5\textwidth}\tiny
    \[\xymatrix{ %3
 1 \ar@{-}[r] & n_1 \ar@{=}[r]     & n_2 \ar@{-}[dr] \ar@{-}[r] & n_3 \ar@{=}[r] & n_4 \ar@{-}[dl] \ar@{-}[r] & n_5 \ar@{=}[r] 
																						& n_6 \ar@{-}[dddlll] \\
              &                    &                            & v_1 \ar@{=}[d] &                            & & \\
			   &                    &                            & v_2 \ar@{-}[d] &                            & & \\
			   &                    &                            & v_3 \ar@{=}[d] &                            & & \\
		       &                    &                            & 2              &                            & & 
}\]
  \end{minipage}
    \caption{Soliton graphs $G_1$, $G_2$ and $G_3$}
    \label{connected:G}
\end{figure}
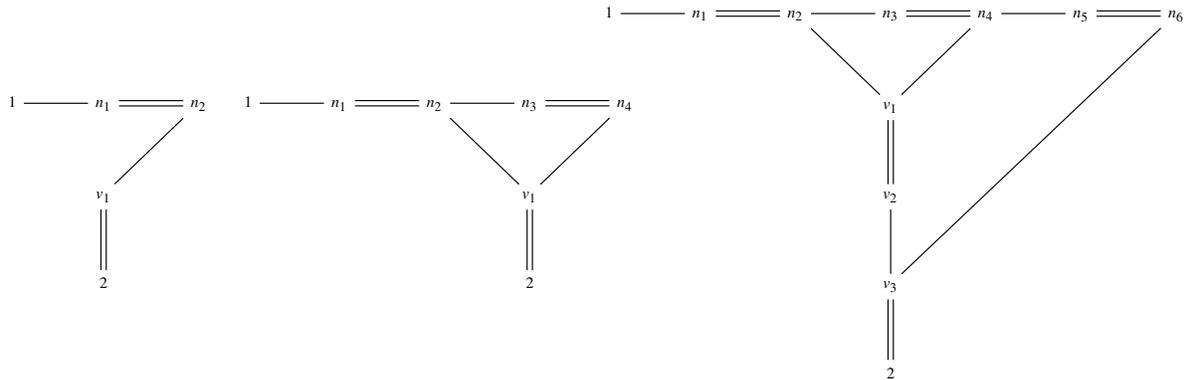

\emph{Proof.} For $g\ge 1$, let~$G_g=(N_g,E_g,w_g)$ be the soliton graph with exactly two exterior nodes~1 and~2 and a path $1,n_1,n_2,\ldots,n_{2g-1},n_{2g}$ with $w_g(1,n_1)=1$ which we will call \emph{basic chain} in the sequel. Moreover,
additional edges leave the basic chain at every other node of the basic chain:
{\small\[\xymatrix{ %g
 1 \ar@{-}[r] & n_1 \ar@{=}[r]     & n_2 \ar@{-}[d] \ar@{-}[r] & n_3 \ar@{=}[r] & n_4 \ar@{-}[d] & \cdots & n_{2g-1} \ar@{=}[r]  & n_{2g} \ar@{-}[d]\\
              &                    &                            &                &                 &        &          &  
}\]}
The edges leaving the basic chain all lead to the exterior node~2 and belong to a sub-graph forming a binary tree with $n_2,n_4,\ldots,n_{2g}$ as leaves and node~2 as its root, in which the root has weight~2 and branching edges always have weight~1. 
The inner nodes of that tree are denoted by $v_1,\ldots,v_r$ with $r=2g-3$ (if $g>1$). There is an edge $(n_2,v_1)$ with weight~1 and, for $1< k\le g$, there is an edge $(n_{2k},v_{2k-3)}$ with weight~1.
The first three soliton graphs $G_1,G_2,G_3$ are depicted in Figure~\ref{connected:G}.

Further, let $B=\{(1,2)\bot\}$. 
 Since the soliton has a non-deterministic choice only on every node $n_{2k}$ of the basic chain, $1\le k< g$, there are exactly~$g$ soliton paths for the soliton in~$B$. Notice that once the soliton has left the basic chain it has to follow the path leading directly to node~2 since it enters a node with degree~3 only when it has just traversed an edge with weight~1. Thus,
 $\vert\result(G_g,B)\rvert\le g$. The soliton uses exactly one of the edges $(n_2,v_1)$ or $(n_{2k},v_{2k-3})$, $1<k\le g$, and the weight of the used edge has turned to~2 whereas the other edges leaving the basic chain keep their weight~1. 
Consequently, $\vert\result(G_g,B)\rvert=g$.

Next, we prove $\result(G',B)=\{G_g\}$ for all $G'\in\result(G_g,B)$ and every $g\ge 1$.
Assume the soliton has used edge $(n_{2k},v_{2k-3})$ in~$G_g$, for some~$k$, $1< k\le g$ (the case~$k=1$, when it has used $(n_2,v_1)$, is similar). Then, the resulting soliton graph~$G'$ is of the form
{\small\[\xymatrix{ %result
 \cdots\ar@{-}[r] & n_{2k-2}\ar@{-}[ddl]\ar@{=}[r] & n_{2k-1}\ar@{-}[r] & n_{2k}\ar@{-}[r]\ar@{=}[ddddlll] & \cdots \\
 \cdots\ar@{-}[d] &                                &                    &                                &        \\
  v_{2k-5}\ar@{=}[d]  &                            &                    &                                &        \\
  v_{2k-4}\ar@{-}[d]  &                            &                    &                                &        \\
  v_{2k-3}\ar@{-}[d]  &                            &                    &                                &        \\
  v_{2k-2}\ar@{=}[d]  &                            &                    &                                &        \\
  \cdots\ar@{-}[d]\ar@{-}[ur]   &                  &                    &                                &        \\
   2                  &                            &                    &                                &
}
\]}
In~$G'$, every soliton path for burst $(1,2)\bot$ has the prefix $1,n_1,n_2,\ldots,n_{2k},v_{2k-3}$. Now, this path can be continued with~$v_{2k-2},v_{2k-1},\ldots,2$ leading directly to node~2, and the resulting soliton graph is~$G_g$.  
Alternatively, the soliton can use edge $(v_{2k-3},v_{2k-4})$ (or, if $k=1$, edge $(v_1,n_4))$, or it can use $(v_{2k-3},v_{2k-2})$ and later an edge with weight~1 leading back towards the basic chain (which has the same weights as in~$G_g$ now). In any such case, some node~$n_j$ of the basic chain will be reached via an edge with weight~1. Therefore, the soliton has to use the edge $(n_j,n_{j-1})$ next, leading "to the left" in the basic chain. Now, it cannot reach node~2, because every node~$n_\ell$ in the basic chain having degree~3 is reached via an edge with weight~1 and has to be left to~$n_{\ell-1}$ since $w(n_\ell,n_{\ell-1})=2$. Therefore, no further soliton graphs are added to $\result(G',B)$.

In conclusion, $\vert\result(G',B)\rvert=1$ for all $G'\in\result(G_g,B)$ and~$g$ is the smallest integer with  $\vert\result(G,B)\rvert\le g$ for all $G\in\states(G_g,B)$. Hence, the degree
of nondeterminism of $\mathcal{A}_B(G_g)$ is~$g$, for every $g\ge1$.
\qed

\section{Graph Properties and Determinism}
\label{sec:Graphs}

So far, we defined determinism properties only on soliton automata.
We now extend our definitions to soliton graphs. 

\begin{definition}[Graph Determinism]\label{def:graphDet}
Let~$G$ be a soliton graph.
$G$ is called 
\begin{itemize}
    \item[(I)] \emph{deterministic}, if for all sets~$B$ of bursts for~$G$ $\mathcal{A}_B(G)$ is deterministic.
    \item[(II)] \emph{strongly deterministic}, if for all sets~$B$ of bursts for~$G$ $\mathcal{A}_B(G)$ is strongly deterministic.
    \item[(III)]  \emph{perfectly deterministic}, if for all sets~$B$ of bursts for~$G$ $\mathcal{A}_B(G)$ is perfectly deterministic.
\end{itemize}
\end{definition}

For our statements about graph determinism it is important to consider soliton graphs that can not be decomposed into independent sub-graphs.
In the case of a single wave, meaning the case of a single soliton traversing a soliton graph, impervious paths may appear.
A path is impervious if none of its edges is used by the soliton in any configuration trail \cite{DassowJurg:Soliton}.
An example of an impervious path is the path h-i-j-k in Figure \ref{fig:1}.
The soliton has to enter the soliton graph either via node~$1$ or node~$2$, hence by traversing an edge with weight~$1$.
Since it has to use an edge with weight~$2$ next, it can only move towards the cycle on the respective side.
On its way back, on node~h or~k, respectively, it has to traverse an edge with weight~$2$ in the next step, still not allowing it to enter the path h-i-j-k and forcing it to leave the soliton graph via the node it entered the graph through.
In order to formalize this idea for the case of multiple solitons being present we give the following definitions.

\begin{definition}[Used Edge]\label{def:usedEdge}
Let $G_0 = (N, E, w)$ be a soliton graph, let $n, n' \in N$ and let~$b$ be a burst.
The edge $(n, n')$ is said to be \emph{used} in a configuration trail $(G_0, \pi_0), (G_1, \pi_1), ..., (G_k, \pi_k)$ with $\pi_0 = \pi_b$ if there exists a soliton $i$ and a timestep $j$ with $0 < j \leq k$, $\pi_{j-1}(i) = n$ and $\pi_j(i) = n'$.
\end{definition}

\begin{definition}[Impervious Path]\label{def:imperv}
Let $G = (N, E, w)$ be a soliton graph and let $n, n' \in N$. A path from~$n$ to~$n'$ is said to be \emph{impervious} if none of its edges are used in a  configuration trail in any~$G' \in \states(G,B)$ with any set of bursts~$B$ for~$G$.
\end{definition}

For the case of a single wave, if a soliton graph contains impervious paths then it can be decomposed into multiple connected components.
Soliton graphs which, after the removal of impervious paths, are connected, are called indecomposable.
For more details see \cite{DassowJurg:Soliton}.

\begin{definition}[Indecomposable Soliton Graph]\label{def:indecomp}
Let $G$ be a soliton graph. $G$ is said to be \emph{indecomposable} if it does not contain an impervious path.
\end{definition}

\begin{definition}[Chestnut]
\label{def:chestnut}
An indecomposable soliton graph is called a \emph{chestnut} if it consists of a single cycle of even length and some paths leading into it with the following conditions: 
\begin{enumerate}
    \item[(I)] Entry points of different paths entering the cycle have even distances;
    \item[(II)] Paths leading to the cycle may meet only at even distances from entry into the cycle.
\end{enumerate}
\end{definition}
\noindent
For more details see \cite{DassowJurg:Soliton}.

\begin{prop}
\label{prop:chestnut}
Let $G=(N,E,w)$ be an indecomposable soliton graph. If~$G$ is a chestnut, then it is not strongly deterministic.
\end{prop}

\begin{figure}
   \begin{minipage}{0.48\textwidth}
     \centering
     \includegraphics[width=.7\linewidth]{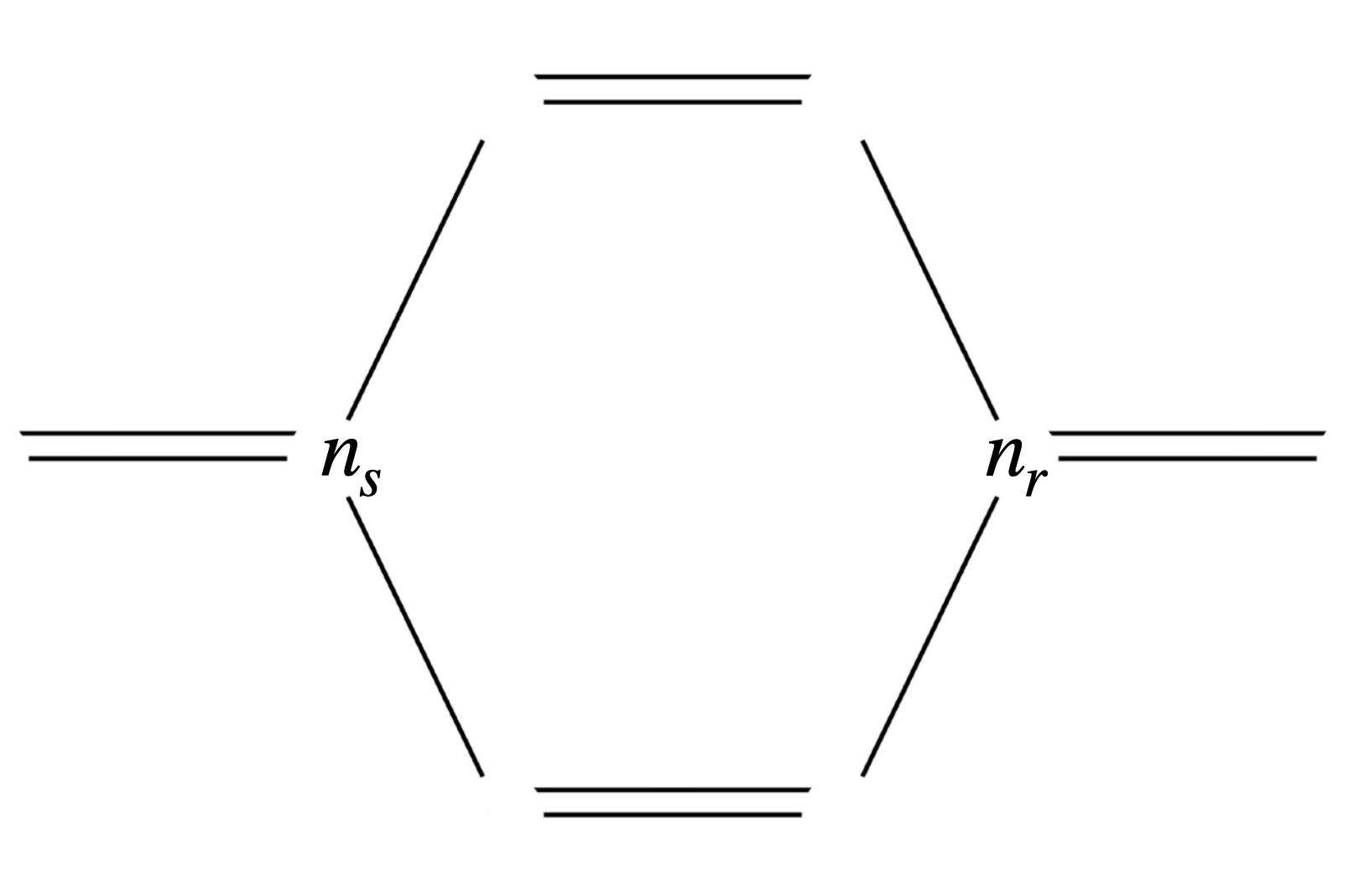}
     \caption{A cycle with even length and two edges with weight~2 leading into it.}\label{fig:3}
   \end{minipage}\hfill
   \begin{minipage}{0.48\textwidth}
     \centering
     \includegraphics[width=.7\linewidth]{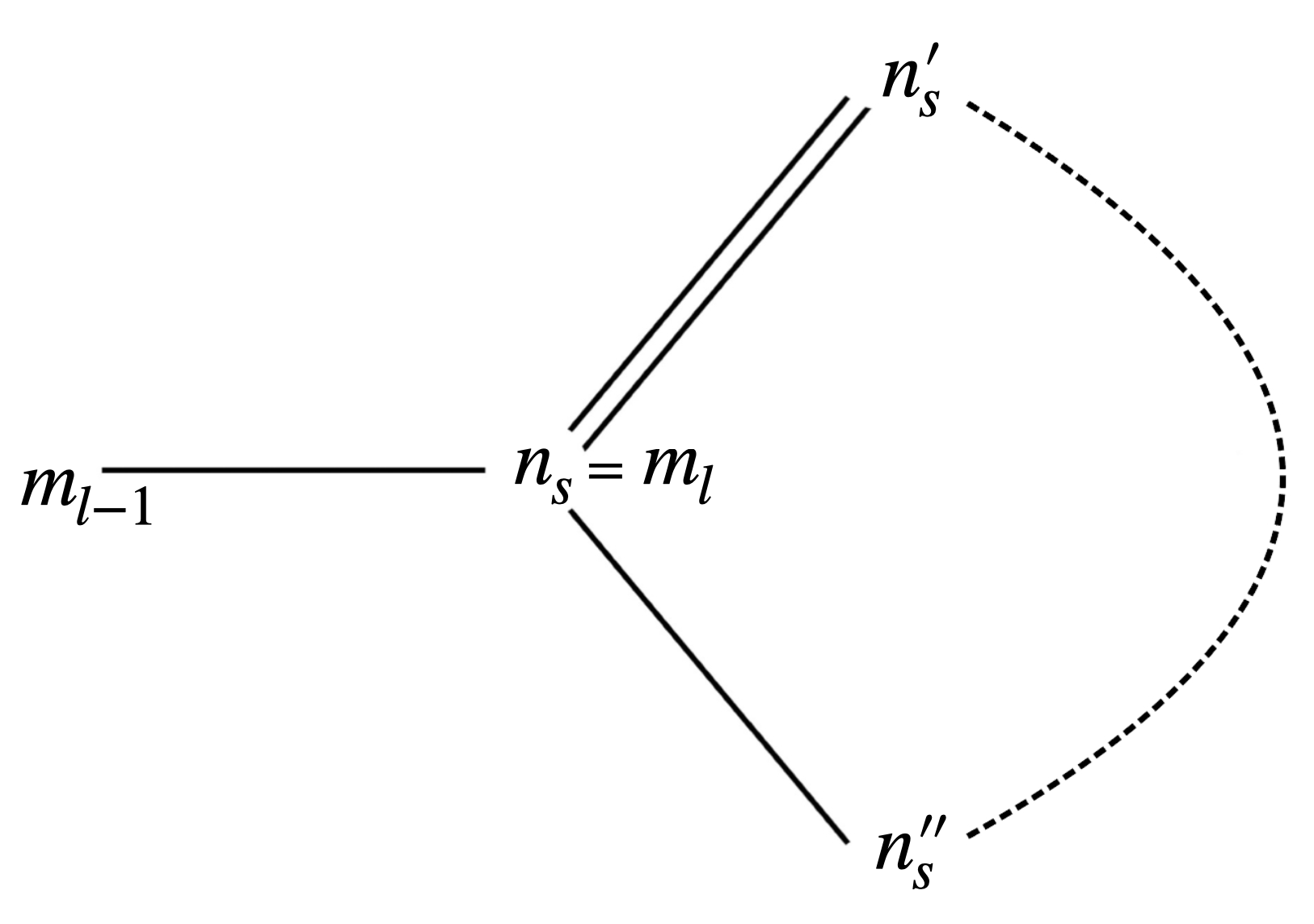}
     \caption{A node of degree $3$ as entry point of a cycle.}\label{fig:4}
   \end{minipage}
\end{figure}

\emph{Proof.} As $G$ is a chestnut, the graph $(N,E)$ contains a cycle of even length (at least $4$), that is there is an integer $k\ge 2$ and a path
$$n_0,\, n_1,\,\ldots,\,n_{2k}$$
with $n_0=n_{2k}$ and $n_i\not= n_j$ for $0\le i<j< 2k$.
For every exterior node~$e$, there is a path leading to the cycle. Without loss of generality, let $m_0,\,m_1,\,\ldots m_\ell$ be such path with $\ell\ge 1$, $m_0=e$ and $m_\ell=n_s$ for some~$s$, $0\le s< 2k$. If $w(m_{\ell -1}, m_\ell) = 2$, then 
$w(n_s,n_{s'}) = w(n_s,n_{s''}) = 1$, where $s'= (s+1)\mod 2k$ and $s''=(s-1)\mod 2k$.
As the length of the cycle is even and two edges with weight~$2$ cannot meet in a soliton graph, there is a node $n_r\not= n_s$ in the cycle such that
$w(n_r,n_{r'}) = w(n_r,n_{r''}) = 1$, $r'= (r+1)\mod 2k$ and $r''=(r-1)\mod 2k$.
An example graph with these properties is visualized in Figure~\ref{fig:3}.
Because~$G$ is a soliton graph, $d(n_r)=3$ and $|r-s|$ is odd.
Thus,~$G$ is not a chestnut. Hence, $w(m_{\ell -1}, m_\ell) = 1$.
Consequently, without loss of generality,  $w(n_s,n_{s'})=2$ and $w(n_s,n_{s''})=1$ (Figure~\ref{fig:4}) and there is a total legal configuration trail for the burst $(e,e)\bot$.
This is seen as follows: the soliton enters the cycle via edge $(m_{l-1}, n_s)$ and changing its weight to~$2$.
It has to continue to $n_s'$.
After completing the cycle it has moved from $n_s''$ to $n_s$ via an edge with weight~$1$ and must leave the cycle to $m_{l-1}$.

Now, consider the burst $b=(e,e)\|_1(e,e)\bot$.
After the first soliton from~$b$ has reached $n_{s'}$, the second one is at node~$n_s$ and must follow the first soliton to~$n_{s'}$, since otherwise the two solitons would collide inside the cycle eventually. When the first soliton has reached~$n_{s}$ again, it must continue to~$n_{s'}$ because it has traversed $(n_{s''},n_s)$ with weight~$1$ and $w(n_s, m_{\ell-1}) = 1$. In the next step (when the second soliton is at~$n_s$), the second soliton has the option to leave the cycle to~$m_{\ell-1}$ or to further follow the first soliton in the cycle. After completing another round through the cycle, exactly the same situation will be encountered again. 
This situation is depicted in Figure \ref{fig:2}, configuration~$11$.

Whenever the second soliton behaved to leave the cycle, the first soliton will be able to complete its path to~$n_{s}$ and then also leave the cycle from~$n_s$ to $m_{\ell-1}$ and on to~$e$. In conclusion, there is more than one total legal configuration trail for~$b$. Hence,~$G$ is not strongly deterministic. \qed

Looking at the details in this proof, several (an infinite number of) imperfect configuration trails, but only one perfect configuration trail, exist.
That is why one might wonder, whether all chestnuts are perfectly deterministic.
The following statement disproves this assumption.

\begin{prop}
There is a chestnut~$G$ which is not perfectly deterministic.
\end{prop}

\begin{figure}
	\centering
	\includegraphics[width=0.98\linewidth]{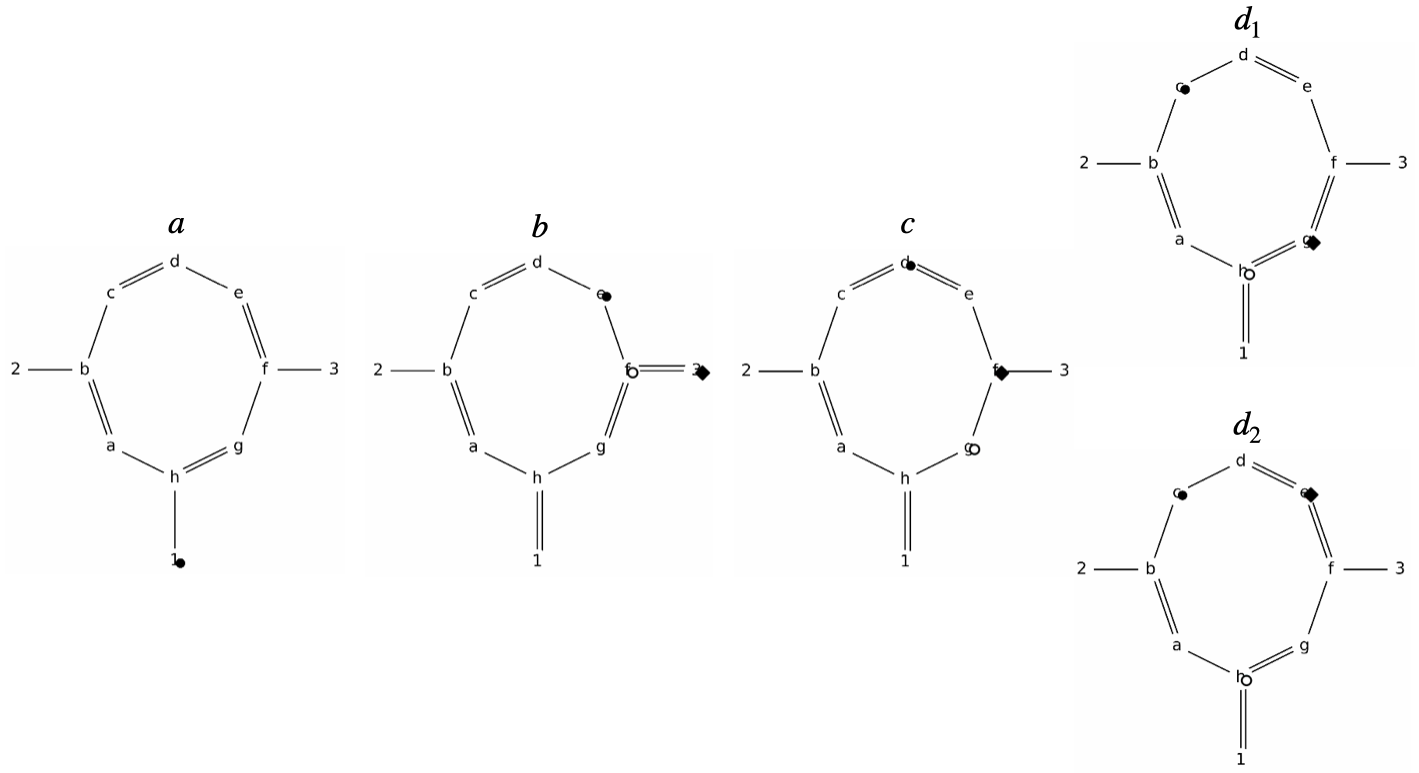}
	\caption{Selected configurations of two configuration trails for the burst $(1,1)\|_3(3,1)\|_1(3,1)\bot$. The first soliton is depicted as a black pebble, the second soliton as a white pebble and the third soliton as a black diamond. Configurations~$a$,~$b$ and~$c$ appear in both configuration trails, while~$d_1$ is part of the first trail and~$d_2$ is part of the second trail.}
	\label{fig:5}
\end{figure}

\textit{Proof.} 
Let~$G$ be the chestnut in configuration~$a$ in Figure~\ref{fig:5} and let $b_G = (1,1)\|_3(3,1)\|_1(3,1)\bot$ be a burst.
There are two total legal configuration trails for~$G$ and~$b_G$.
We show selected configurations of both trails in Figure~\ref{fig:5}, which are $a$, $b$, $c$, $d_1$ for the first and $a$, $b$, $c$, $d_2$ for the second trail.
They differ in the third soliton, depicted as a black diamond, moving downwards to node~$g$ in $d_1$ and upwards to node~$e$ in $d_2$.
Therefore, both configurations trails are perfect. As both can be continued to total legal configuration trails,~$G$ is not perfectly deterministic. 
\qed

\begin{prop}
\label{prop:tree}
Let $G = (N, E, w)$ be an indecomposable soliton graph.
If $(N, E)$ is a tree, then~$G$ is strongly deterministic.
\end{prop}

\textit{Proof.} 
Let $G=(N,E,w)$ be an indecomposable soliton graph, $X$ be the set of its exterior nodes and~$B$ a set of bursts over $X\times X$. 
Let $(n,n')$ be any pair of exterior nodes. If $(N,E)$ is a tree, then there is exactly one path $n_0,n_1,\ldots n_k$ from~$n_0=n$ to~$n_k=n'$ such that $i\not=j$ implies $n_i\not= n_j$, that is, no node occurs more than once. In every total legal configuration trail~$C$ for~$G$ and a burst~$b\in B$, condition 2.\textit{(c)v.}\ of Definition~\ref{def:configTrail} guarantees that only paths with that property can be a soliton path of a soliton in~$b$ (solitons are not allowed to turn around spontaneously).
Consequently, for every
soliton~$i$ in $b\in B$ and every total legal configuration trail~$C$ for~$G$ and~$b$, there is at most one soliton path of soliton~$i$ in~$C$. 
Therefore, the automaton $\mathcal{A}_B(G)$ is strongly deterministic. As~$B$ was arbitrary,~$G$ is strongly deterministic.
\qed

\begin{theorem}
Let $G = (N, E, w)$ be an indecomposable soliton graph.
$G$ is strongly deterministic if and only if $(N, E)$ is a tree.
\end{theorem}

\emph{Proof.}
For single-soliton automata~(\cite{DassowJurg:Soliton}) it is known that an indecomposable solition graph is strongly deterministic if and only if~$G$ is a chestnut or $(N, E)$ is a tree, see Proposition~5.4 in~\cite{DassowJurg:Soliton}. Proposition~31 of~\cite{BorJur:Multiwave} implies for every soliton graph~$G$ with set~$X$ of exterior nodes that the single-soliton automaton based on~$G$ is the soliton automaton~$\mathcal{A}_B(G)$ where $B=\{\,(n,n')\bot\mid n,n'\in X\,\}$.\footnote{See also Definition~10 in~\cite{BorSchu:Determinism}.} In conclusion, if~$G$ is neither a chestnut nor is $(N,E)$ a tree, then there is a set of bursts~$B$ such that~$\mathcal{A}_B(G)$ is not strongly deterministic, thus~$G$ is not strongly deterministic. By Proposition~\ref{prop:chestnut}, $G$ is not strongly deterministic, if it is a chestnut. Therefore, if~$G$ is strongly deterministic, then $(N,E)$ is a tree. By Proposition~\ref{prop:tree}, the statement follows.
\qed

\section{Concluding Remarks}

So far, the restriction for soliton automata to be (strongly) deterministic has only been investigated for the single-soliton case in the literature, see~\cite{DassowJurg:Soliton}. In~\cite{BorSchu:Determinism,Schulz:BSc:2023} several concepts of determinism have been defined for multi-soliton automata, but they have not been further investigated. In the present paper, the new notion of perfect determinism is defined, forming a weaker requirement than strong determinism but a stricter requirement than determinism. A characterization of strongly deterministic soliton graphs is given that is deviating from the known result for single-soliton automata. An example of a soliton graph is presented that is strongly deterministic in the single-soliton case but is not even perfectly deterministic in the multi-soliton case. The degree of non-determinism is shown to be a connected measure of descriptional complexity for soliton automata.

The results use the condition that the soliton graphs are indecomposable, that is, there are no impervious paths in the soliton graphs. An interesting research question is whether impervious paths can appear at all in soliton graphs in the multi-soliton case. A soliton passing a node of a path that is impervious to that soliton opens the path for a second soliton following. The question is whether or not this principle can be generalized to open an unbounded number of impervious paths which may be "hidden" behind each other without eventually causing collisions so that each soliton can leave the graph again, constituting a total legal configuration trail for the respective burst. 

In addition to the characterization of strongly deterministic soliton graphs one could also seek to characterize perfectly deterministic and deterministic soliton graphs.
Another field of future research is the investigation of the transition monoids of multi-soliton automata.

\nocite{*}
\bibliographystyle{eptcs}
\bibliography{references}
\end{document}